\documentclass[10pt,twocolumn]{article}

\usepackage[a4paper,
            top=1.8cm,
            bottom=2.0cm,
            left=1.6cm,
            right=1.6cm,
            columnsep=0.6cm]{geometry}

\usepackage{times}
\usepackage{algorithm}
\usepackage{algpseudocode}
\usepackage{svg}
\usepackage{float}
\usepackage{multirow}
\usepackage{makecell}
\usepackage{microtype}
\usepackage{graphicx}
\usepackage{comment}
\usepackage{booktabs}
\usepackage{amsmath,amssymb}
\usepackage{mathtools}
\usepackage{enumitem}
\usepackage{xcolor}
\usepackage{url}
\usepackage[hidelinks]{hyperref}
\usepackage[numbers,sort&compress]{natbib}
\usepackage{balance}
\usepackage{colortbl}
\usepackage{titlesec}
\usepackage[font=small,labelfont=bf]{caption}
\usepackage{amsthm}

\titleformat{\section}
  {\normalfont\bfseries\large}
  {\thesection}{0.6em}{}
\titleformat{\subsection}
  {\normalfont\bfseries\normalsize}
  {\thesubsection}{0.6em}{}
\titleformat{\subsubsection}
  {\normalfont\bfseries\normalsize}
  {\thesubsubsection}{0.6em}{}

\titlespacing*{\section}{0pt}{1.2ex plus 0.3ex minus 0.2ex}{0.6ex}
\titlespacing*{\subsection}{0pt}{1.0ex plus 0.2ex minus 0.2ex}{0.4ex}
\titlespacing*{\subsubsection}{0pt}{0.8ex plus 0.2ex minus 0.2ex}{0.3ex}

\setlist{nosep,leftmargin=*}

\definecolor{lightgrayrule}{gray}{0.8}

\title{\bfseries Collective Autonomous Vehicle Routing Needs Social Awareness for Coexistence with Human Drivers}

\author{
Anastasia Psarou$^{*}$,
Mikolaj Rams,
Blazej Torbus,
Rafal Kucharski \\[0.5em]
Faculty of Mathematics and Computer Science, Jagiellonian University \\
Kraków, Poland
}

\date{}

\begin{document}

\twocolumn[
\begin{@twocolumnfalse}
\maketitle

\begin{abstract}
\normalsize
Collective Autonomous Vehicle (AV) route choices in real-world traffic networks can create externalities for Human-Driven Vehicles (HDVs), as routes that benefit AVs may increase travel times for other road users.
To address this, we introduce a social component, the marginal cost, to the selfish travel time AV reward. Marginal cost captures an AV's impact on other road users by comparing their travel times with those in a counterfactual simulation in which the AV is removed.

We study AV route choice in mixed-autonomy traffic settings using Multi-Agent Reinforcement Learning (MARL), where AVs learn routing policies through repeated interactions with the traffic environment.
Experiments on the real-world networks of Ingolstadt and Provins show that \textbf{incorporating marginal cost into the AV reward not only limits the increase in HDV travel times and shortens AV travel times relative to selfish AV routing, but also improves overall system performance}.
In Ingolstadt, selfish routing of 40\% AVs increases the travel times of other vehicles by 1.7\%, and this drops to 1.1\% when the social reward is included.
Interestingly, not only does overall system performance improve, but the AVs are also better off when acting socially.
This effect is consistent across tested cities, penetration rates, and replications, yet when the weight of the social component in the reward increases, the system fails to reach the desired performance.

For a more scalable approach to social AV routing in mixed-autonomy traffic, we compute the marginal cost only for the AVs whose route choices most affect the traffic system. This reduces the number of counterfactual simulations and achieves similar performance at a significantly lower computational cost.

\textbf{Our results suggest that AV routing should go beyond selfish rewards and, for the benefit of all traffic participants, include a social component in its objective.}
\end{abstract}

\vspace{0.5em}

\noindent\textbf{Keywords:}
Autonomous vehicles, urban route choice, multi-agent reinforcement learning, social awareness, reward shaping.

\vspace{2.0em}
\end{@twocolumnfalse}
]

\footnotetext{${}^{*}$Corresponding author: \texttt{anastasia.psarou@uj.edu.pl}}

\section{Introduction}

Every day, human drivers commute through urban traffic networks, choosing among several possible routes the ones with the minimal expected perceived travel time \cite{WARDROP1952,inductive_bias_bounded_rationality}. The growing presence of Autonomous Vehicles (AVs) in cities may introduce new dynamics, as AVs may deviate from the shortest routes provided by real-time navigation systems and can use a variety of algorithms to learn routing strategies, potentially achieving shorter travel times than Human-Driven Vehicles (HDVs) or disadvantaging them by consistently increasing their travel times \cite{jamroz2025social, psarou2025collaborationcitymachinelearning}.

\begin{figure*}[t]
    \centering
    \includegraphics[
        width=\textwidth,
        trim={3.5cm 2.5cm 3.5cm 1.2cm},
        clip
    ]{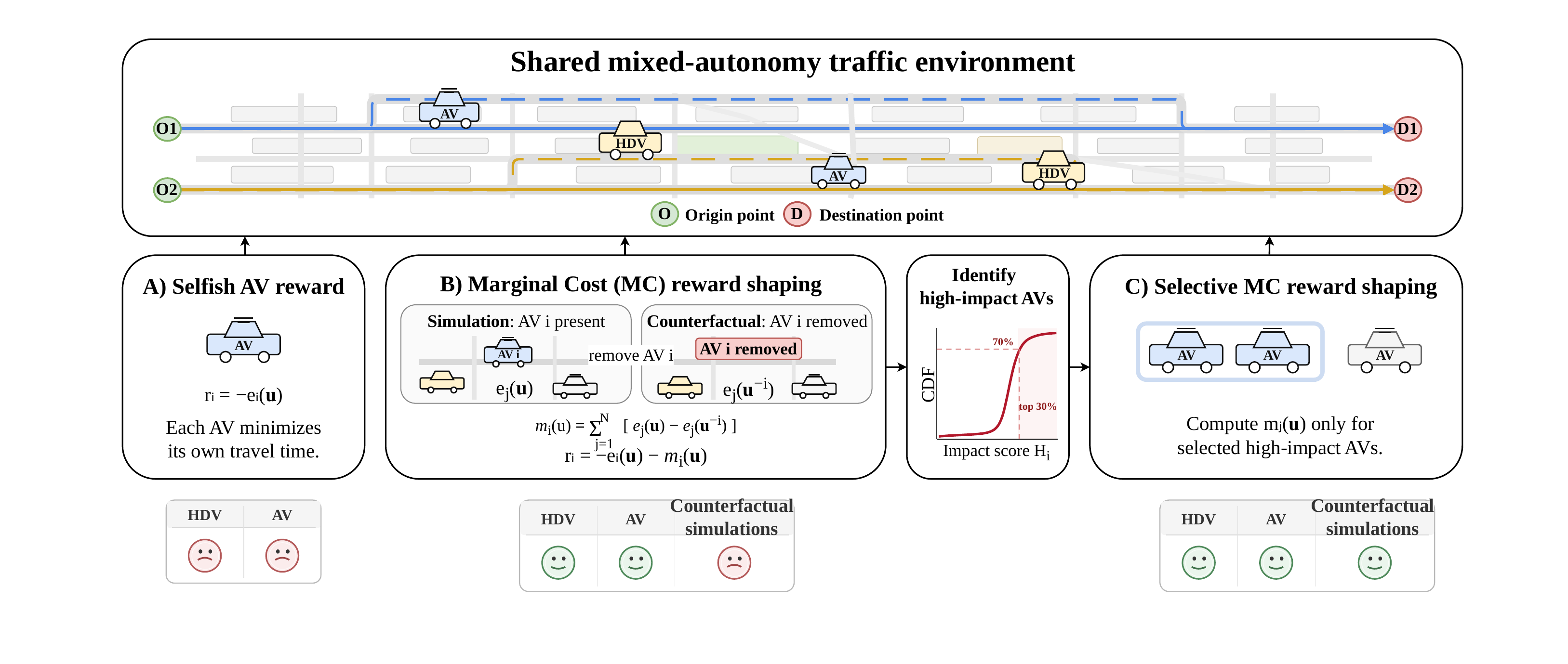}
    \caption{We study AV route choice in traffic environments shared with human drivers. (A) When AVs selfishly minimize their own travel times, their routing decisions can increase HDV travel times. (B) Incorporating our marginal cost into the AV reward allows AVs to account for the impact of their route choices on other road users, mitigating the increase in HDV travel times observed under selfish AV routing while also improving AV travel times on average. However, computing marginal cost requires additional counterfactual simulations for each AV, adding computational cost. (C) To reduce this burden, we introduce selective marginal cost reward shaping, which computes marginal cost rewards only for a subset of the most impactful AVs. This preserves some of the benefits for AVs and HDVs' travel times while reducing the number of counterfactual simulations.}
    \label{fig:overview}
\end{figure*}

We consider a city that initially consists entirely of HDVs who learn routing strategies by updating their utility through repeated days of commuting \cite{Cascetta2009}. We assume this repeated learning process brings the traffic system close to an equilibrium state \citep{WATLING2018254}. Then, a subset of HDVs switches to traveling with AVs, which use Multi-Agent Reinforcement Learning (MARL) to learn routing policies. As learning progresses, AVs reduce their travel times and eventually achieve shorter average travel times than HDVs. However, we find that when AVs select their routes solely to minimize their individual travel times (selfish behavior), their routing policies can increase the travel times of HDVs relative to those experienced before AVs were introduced in the system (see Figure \ref{fig:overview}A).

We therefore ask: \textbf{How can AV routing MARL algorithms be designed to reduce AV travel times while limiting the additional costs on other road users, given that urban road networks are shared public infrastructure?} To solve this problem, we augment the selfish travel-time AV reward with a component that accounts for each AV's impact on the system, the marginal cost, following the formulation introduced in \cite{10.65109/FTHN6981}. We show that incorporating marginal cost into AV rewards can reduce the increase in HDV travel times and allow AVs to achieve shorter travel times than under selfish AV routing (see Figure \ref{fig:overview}B). We further find that these travel-time improvements depend on the relative weighting of the two reward components: assigning a greater weight to marginal cost can further improve HDV travel times but may increase AV and overall system travel times.


We further investigate how to make marginal cost computation scalable, since it requires an additional counterfactual simulation per AV each day of commute. In each counterfactual, we remove an AV from the traffic system and compare the remaining vehicles' travel times with those observed when the AV is present. For example, with 400 AVs and 10,000 training iterations (equivalent to commute days), computing marginal costs for all AVs would require four million additional simulation runs. We therefore investigate whether marginal cost rewards can be computed and applied only to a fraction of the most impactful AVs on the traffic system (selective marginal cost reward shaping), rather than to all AVs (marginal cost reward shaping). We then evaluate whether selective marginal cost reward shaping can achieve system, AV, and HDV travel-time improvements comparable to those obtained with marginal cost reward shaping, relative to selfish AV routing (Figure \ref{fig:overview}C).


The contributions of this work include:

\begin{itemize}[topsep=0pt]
     \item We show that AV routing policies trained with MARL to minimize individual AV travel times can increase HDV travel times relative to those experienced before the introduction of AVs in the system, revealing a negative externality of selfish AV routing (see Figure \ref{fig:relative_travel_time_improvement}A, D).
     \item We demonstrate that incorporating a social component, the marginal cost, into the AV reward can mitigate the increase in HDV travel times observed after AV introduction while also improving AV travel times compared to those achieved when AVs minimize their individual travel times. We further show that AV and HDV travel times depend on how much weight is assigned to the marginal cost. Higher weights can further improve HDV travel times but worsen AV and overall system performance (see Figures \ref{fig:relative_travel_time_improvement}, \ref{fig:human_av_tt_ratio}).
    
    \item We show that computing marginal costs only for a fraction of the most impactful AVs, \textit{selective marginal cost reward shaping}, is sufficient to retain HDV, AV, and system travel times comparable to those achieved when all AVs receive the marginal cost reward. This approach also reduces the number of counterfactual simulations (see Table \ref{tab:selective_mc_example}). We identify impactful AVs from sampled joint actions that capture how total system travel time varies across alternative AV route choices.
\end{itemize}

\subsection{Literature review}
\label{sec:related_works}

Route choice can be formulated as a congestion game in which travelers repeatedly select among alternative routes in a traffic network while sharing limited road capacity \citep{holzman1997strong}. Assuming homogeneous, fully rational users, this game converges to the Wardrop or User Equilibrium (UE), the Nash equilibrium in the route choice context, where no traveler can improve their travel time by unilaterally switching routes \citep{Wardrop1952ROADPS, Beckmann1956}. The System Optimum (SO) is the traffic allocation that minimizes total system travel time across the network \citep{merchant1978optimality}, but achieving it may require some travelers to take routes that are less favorable for them to improve overall system performance.

Human route choice is commonly modeled using the utility-maximization decision rule \cite{DUNCAN2025, jamroz2025social, LI2023103627}. The main component of the utility is the expected travel time for each route \cite{doi:10.3141/2014-01}, which is updated after each trip. In random utility theory, utility additionally includes a random stochastic variable that captures heterogeneity in travelers' preferences, imperfect perception of travel time and congestion, and fluctuations in driving conditions \cite{Cascetta2009, WATLING2018254}. This stochastic route choice behavior is assumed to converge to the Stochastic User Equilibrium (SUE), in which no driver \textit{believes} they can improve their travel times by switching routes.

In a mixed-traffic environment where HDVs and AVs coexist, AV route assignment can be solved in several ways. MARL is a promising method \cite{akman2024impact, akman2025urburbanrouting, AgentRewardShaping, lazar2021learningdynamicallyrouteautonomous, Thomasini+2023}, where AVs can learn routing policies through repeated interactions with the traffic environment. These policies can be optimized using different objectives, such as minimizing AVs' individual travel time, the system's travel time, or even maximizing HDV travel times \cite{akman2024impact, jamroz2025social, akman2025urburbanrouting}. Among these objectives, if AVs optimize only their individual travel times, their routing choices may increase other road users' travel times because of limited road capacity. This motivates the need to search for AV routing objectives that improve AV travel times while limiting the additional costs imposed on other road users.

In economics, marginal cost denotes the change in total cost resulting from an increase in the quantity produced \cite{mas1995microeconomic}. In transportation research, marginal cost quantifies a vehicle's impact on the system by measuring the additional travel time it imposes on other road users through its presence in the traffic system \cite{peeta1995system, sheffi1985urban}. This idea forms the basis of marginal cost pricing, in which travelers are charged according to the increase in latency their action or presence causes to the system \cite{COLE2006444, 10.2307/2224491}. Marginal cost pricing has also been incorporated into reinforcement learning approaches for route choice. Ramos et al. \cite{toll_based_learning_minimizing_congestion}, for example, incorporated marginal cost tolls into the reward signal for commuters using reinforcement learning for route choice, aiming to align individual routing decisions with system-level efficiency and minimize congestion. Inspired by the concept of marginal cost and following \cite{10.65109/FTHN6981}, we define the marginal cost of an AV as the difference in travel time for all vehicles in a system when the AV is present versus a counterfactual simulation in which that AV is removed.

Marginal cost relates to difference rewards, introduced in cooperative multi-agent settings to address credit assignment when agents optimize a shared global objective \cite{AgentRewardShaping, 2615731.2615761, 2343896.2344025}. Difference rewards compare the global reward with a counterfactual in which an agent’s contribution is removed or replaced by a default action, providing a reward that reflects the agent’s individual contribution to the system. In traffic routing, Grunitzki et al. \cite{grunitzki2014individual} studied route choice with reinforcement learning and compared agents trained with selfish travel time rewards and difference rewards, showing that the latter can achieve lower average system travel times.

Prior work introduced the marginal cost reward to help MARL routing AVs reach SO solutions faster than selfish routing AVs in a toy network where the UE and SO solutions coincided \cite{10.65109/FTHN6981}. In this work, we examine whether marginal cost reward shaping can prevent selfish AV routing from reducing AV travel times at the expense of HDVs in real-world traffic networks. We additionally propose a method to decrease the counterfactual simulations needed to compute the marginal cost reward by rewarding a fraction of the most impactful AVs in the system.

In the literature, researchers have studied the identification of impactful agents to explain agents' importance in multi-agent systems. For example, Heuillet et al., \cite{10.1109/MCI.2021.3129959} used Shapley values to quantify each agent's contribution to the global reward. Chen et al. \cite{10.1609/aaai.v39i15.33733} stated that an agent is impactful if the system total reward is largely changing when its action is randomized. Following a similar intuition, we consider an AV to be impactful when alternative route choices lead to substantial differences in total system travel time.

\textbf{Research gap:} To the best of our knowledge, little attention has been given to designing AV routing objectives that minimize AV travel times while mitigating travel time increases for other road users in mixed-autonomy traffic. Prior work has explored different AV routing objectives but has not explicitly investigated how to mitigate the travel time externalities these objectives may impose on HDVs or the system \cite{akman2024impact, jamroz2025social, akman2025urburbanrouting}. Similarly, existing work on difference rewards has primarily focused on cooperative settings, whereas we use the counterfactual term to account for the externalities that one agent population can impose on another while coexisting in the same environment \cite{AgentRewardShaping, 2615731.2615761, 2343896.2344025, grunitzki2014individual}.

We address this gap by incorporating marginal cost into the AV routing reward and examining how the trade-off between individual AV travel times and their system impact during MARL training affects AV, HDV, and system travel times after training. Since AV routing objectives can shape how congestion costs are distributed among agents participating in this complex system. We further investigate selective marginal cost reward shaping, in which we apply marginal cost only to a fraction of the most impactful AVs. We examine whether we can achieve similar AV, HDV, and system travel times with fewer counterfactual simulations, providing a more scalable approach to social AV routing.

\section{Preliminaries}

\subsection{Urban routing problem}
\label{sec:simulation}

We study the route choice problem in which, each day, a number of vehicles make one trip from an origin to a destination point in an urban traffic network at a specific departure time by selecting a route from a set of discrete alternatives. To simulate this problem, we use the \textit{RouteRL} \cite{akman2025routerlmultiagentreinforcementlearning} framework integrated with \textit{SUMO} \cite{SUMO2018}, a continuous microscopic traffic simulator that enables modeling complex urban traffic networks and agent interactions.

Our urban environment initially consists only of HDVs, which choose routes using random utility theory \cite{Cascetta2009}. After consecutive days of commuting, HDV route choices stabilize. Then, a number of HDVs switch to using AVs, which can use various MARL algorithms to learn routing strategies. AVs keep the departure time, origin, destination points, and action space (set of routes they can travel on) of their human predecessors. Human learning is thereafter disabled, and human route choices are sampled from the fixed learned distribution. This provides behaviorally plausible human route choices that AV routing policies can be trained on before deployment. AV observations include an estimate of each route's travel time based on travel times of earlier-departing vehicles with the same origin and destination, together with the AV's origin, destination, and departure time. The set of routes does not change during the simulation or from one day to the next, and both departure times and demand remain constant across simulation days.

\subsection{Problem formulation}
\label{sec:problem_formulation}

We abstract the daily (repeated) route choices made by humans (and AVs in the future) in capacitated networks (limited resources) as a repeated congestion game \cite{holzman1997strong}. This can be formalized as a one-cycle Agent Environment Cycle (AEC) problem \cite{terry2021pettingzoogymmultiagentreinforcement}, defined in \cite{akman2025urburbanrouting, psarou2025collaborationcitymachinelearning} as a tuple:

\begin{equation*}
\left\langle 
   \mathcal{S}, \mathcal{N}, \mathcal{I},
   \{\mathcal{U}_i\}_{i \in \mathcal{N}},
   \{r_i\}_{i \in \mathcal{I}},
   \{\Omega_i\}_{i \in \mathcal{I}},
   \upsilon,
\right\rangle
\end{equation*}

where $\mathcal{N}=\mathcal{I} \cup \mathcal{H}$ denotes the set of all agents, with $\mathcal{I}$ representing AV agents and $\mathcal{H}$ human agents. In each episode (day or iteration), each AV agent in $\mathcal{I}$ and each human agent in $\mathcal{H}$ selects a route from their respective action space, with agents acting sequentially according to $\upsilon$. The AEC game tuple includes:

\begin{itemize}
\item A finite set of states $\mathcal{S}$.
\item  A finite set of AV agents $\mathcal{I}$, a finite set of human agents $\mathcal{H}$.
\item For each agent $i \in \mathcal{N}$:
\begin{itemize}
    \item The agent selects an action $u_i \in U_i$, forming a joint action $\mathbf{u} \in \mathbf{U} \;\equiv\; U^N$.
\end{itemize}
\item For each agent $i \in \mathcal{I}$:
\begin{itemize}
    \item $r_i$ is the immediate reward agent $i$ aims to maximize, as defined in \cite{akman2025urburbanrouting, psarou2025collaborationcitymachinelearning}.
    \item $\Omega_{i}$ is the set of possible observations for agent $i$.
\end{itemize}

\item $\upsilon$ is the agent-selection mechanism that determines which agent acts next by sorting agents by their departure time from their origin.

\end{itemize}

In this problem, AVs learn routing policies $\pi_i(u_i \mid \omega_i)$ over consecutive days (episodes) simulated in SUMO to maximize their immediate rewards. A wide range of MARL algorithms can be used to learn such policies, which select routes based on each AV's current observation \(\omega_i\).

\section{Method}

\subsection{Reward with marginal cost}
\label{sec:marginal_cost}

We postulate that adding to the AV reward a component that captures the agent's impact on all other agents in the system, in our case, the marginal cost, can guide AVs towards route choices that account for the travel time increases they may impose on HDVs. We formulate the reward of an AV agent $i$ as:

\begin{equation}
r_i(\mathbf{u}) = - e_i(\mathbf{u}) - \beta m_i(\mathbf{u}), \qquad i \in \mathcal{I}
\label{eq:immediate_reward}
\end{equation}

where $e_i(\mathbf{u})$ denotes the travel time experienced by individual AV $i$, under the joint action $\mathbf{u}$ as computed by SUMO, and $m_i(\mathbf{u})$ represents the marginal cost of AV $i$ in the system. The parameter $\beta$ weights the marginal cost component in the agent's reward. The marginal cost of an agent $i$ is represented by:

\begin{equation}
m_i(\mathbf{u}) =
\sum_{\substack{j=1 \\ j \neq i}}^{N}
\!\left( e_j(\mathbf{u}) - e_j(\mathbf{u}^{-i}) \right), \quad  N = |\mathcal{H} \cup \mathcal{I}|,
\label{eq:marginal_cost}
\end{equation}

and corresponds to the travel time differences of all the agents $j$ in the system, when agent $i$ is present, and in a counterfactual simulation where agent $i$ is removed. $\mathbf{u}^{-i}$ denotes the joint action of the system where agent $i$ is not present in the simulation.

When $\beta=0$, AVs behave selfishly by minimizing their individual (selfish) travel time. Positive values of $\beta$ introduce a trade-off between AVs routing selfishly and accounting for the impact of their route choices on the system.

\begin{algorithm}[t]
\caption{AV training with marginal cost reward}
\label{alg:marginal_cost_training}
\begin{algorithmic}[1]
\Require Human agents $\mathcal{H}$, AV agents $\mathcal{I}$,
         AV policies $\{\pi_i\}_{i\in\mathcal{I}}$, 
         HDV route-choice distributions $\{P_h\}_{h\in\mathcal{H}}$,
          set of AVs for which we compute the marginal cost $\mathcal{I}_{\mathrm{MC}} \subseteq \mathcal{I}$,
         coefficient $\beta$

\Statex
\For{each episode (day)}
    \State $\mathbf{u} \gets \emptyset$

    \Statex \texttt{// Main simulation}
    \For{each agent $k \in \mathcal{H}\cup\mathcal{I}$ by $\upsilon$ order} 

        \State Advance the simulation to agent $k$'s decision time

        \If{$k \in \mathcal{H}$} \hspace{0.5em}$\triangleright$ HDVs
            \State Sample route $u_k \sim P_k
        $
        \Else \hspace{0.5em}$\triangleright$ AVs
            \State Observe $\omega_k$ and select route $u_k \sim \pi_k(\cdot \mid \omega_k)$
        \EndIf

        \State Agent $k$ takes route $u_k$, $\mathbf{u}[k] \leftarrow u_k$
        
    \EndFor
    \State Record travel times
    $\{e_j(\mathbf{u})\}_{j\in\mathcal{H}\cup\mathcal{I}}$
    \Statex

    \Statex \texttt{// Counterfactual simulations}
    \For{each AV $i \in \mathcal{I}_{\mathrm{MC}}$}

        \For{each agent $k \in \mathcal{H}\cup\mathcal{I}$
             in $\upsilon$ order}
            \State Advance the simulation to agent $k$'s departure time

            \If{$k \neq i$}
                \State Apply the stored route $u_k$
            \EndIf
        \EndFor
        
        \State Record counterfactual travel times
        $\{e_j(\mathbf{u}^{-i})\}_{j\neq i}$

        \State $m_i(\mathbf{u}) \gets
        \displaystyle
        \sum_{j\neq i}
        \left(
        e_j(\mathbf{u}) - e_j(\mathbf{u}^{-i})
        \right)$

        \State $r_i(\mathbf{u}) \gets
        -e_i(\mathbf{u})-\beta m_i(\mathbf{u})$
    \EndFor
    \Statex
    \Statex \texttt{// AV learning}
    \For{each AV $i \in \mathcal{I}$}
        \State Update policy $\pi_i$ using $r_i(\mathbf{u})$
    \EndFor
\EndFor
\end{algorithmic}
\end{algorithm}

\subsection{Marginal cost table}

To compute the marginal cost for each AV, we construct the marginal cost matrix $M(\mathbf{u})$, which records the impact of each AV on the travel times of all other agents in the system. Each entry $M_{i, j}(\mathbf{u})$ quantifies the impact of agent $j$ on the travel time of agent $i$. 

\begin{equation}
M_{i,j}(\mathbf{u}) =
\begin{cases}
e_i(\mathbf{u}) - e_i(\mathbf{u}^{-j}), & i \neq j, \\
0, & i = j,
\end{cases}
\qquad j \in \mathcal{I}.
\label{eq:marginal_cost_matrix}
\end{equation}

The marginal cost of agent $j$, $m_j(\mathbf{u})$, is obtained by summing the entries in column $i$ of the marginal cost matrix, capturing the total change in travel time experienced by all agents in the system due to the presence of AV $i$.

We build the marginal cost matrix $M(\mathbf{u})$ by rerunning the simulation for a specific joint action $\mathbf{u}$, once for each AV agent. In each additional simulation, we remove a single AV $j$, while keeping the routes and departure times of all other agents fixed. As a result, constructing the matrix for one episode (joint action) requires generating $|\mathcal{I}|$ additional and separate environment instances. Since these counterfactual simulations are independent, they can be executed in parallel to reduce computation time. Algorithm 1 details the marginal cost reward computation during AV training.

\subsection{Selective marginal cost}
\label{sec:selective_marginal_cost}

To reduce the number of simulations needed to compute marginal cost counterfactuals for all AV agents in the system, we propose selective marginal cost reward shaping. In selective marginal cost reward shaping, we rank AVs by their impact on the system. We assess an AV's impact by how much system travel times differ when it chooses different routes. Then, we compute the marginal cost $m_i$ only for a fraction of the most impactful AVs, forming the subset $\mathcal{I}_{\text{MC}} \subseteq \mathcal{I}$. The remaining AVs aim to minimize their own travel time. In this way, selective marginal cost reward shaping concentrates the additional counterfactual computation on the AVs with the largest estimated system impact, allowing us to investigate whether similar AV, HDV, and system travel times can be achieved with fewer counterfactual simulations.

To identify impactful AV agents, we construct a dataset $\mathcal{D}$ of sampled joint action realizations. Each sample \(d \in \mathcal{D}\) represents a day of commute, where \(\mathbf{u}^{(d)} = (u_1^{(d)},\ldots,u_N^{(d)})\) is the vector of joint actions and the vector of realized travel times is \(\mathbf{e}^{(d)} = (e_1(\mathbf{u}^{(d)}),\ldots,e_N(\mathbf{u}^{(d)}))\) . For each sample, we compute the total system travel time as:

\begin{equation}
T(\mathbf{u}^{(d)}) =
\sum_{i=1}^{N}
e_i\!\left(\mathbf{u}^{(d)}\right), \quad  N = |\mathcal{H} \cup \mathcal{I}|.
\end{equation}

For each AV agent $i \in \mathcal{I}$ and each available action
$a \in \mathcal{U}_i$, we then extract the subset of samples in which
agent $i$ selects action $a$:

\begin{equation}
\mathcal{D}_{i,a}
=
\left\{
d \in \mathcal{D}
\;\middle|\;
u_i^{(d)} = a
\right\}.
\label{eq:agent_action_subset}
\end{equation}

Then, we compute $\overline{T}_{i,a}$ as the average total system travel time over all such samples where agent $i$ chooses action $a$:

\begin{equation}
\overline{T}_{i,a}
=
\frac{1}{|\mathcal{D}_{i,a}|}
\sum_{d \in \mathcal{D}_{i,a}}
T\!\left(\mathbf{u}^{(d)}\right)
\label{eq:action_conditioned_system_tt}
\end{equation}

Then, for each action $a \in \mathcal{U}_i$ we collect the resulting
action-conditioned system travel times $\overline{\mathbf{T}}_i =
\left[
\overline{T}_{i,a_1},
\overline{T}_{i,a_2},
\ldots,
\overline{T}_{i,a_{|\mathcal{U}_i|}}
\right]$.
We then quantify the historical impact of agent $i$, denoted as $H_i$, as the standard
deviation of the average system travel times across each possible action. 

\begin{equation}
H_i =
\operatorname{Std}_{a \in \mathcal{U}_i}
\left(
\overline{T}_{i,a}
\right).
\label{eq:historical_agent_impact}
\end{equation}

A high $H_i$ value indicates that, among the samples in dataset $\mathcal{D}$, the system travel times differ substantially when agent $i$ chooses different routes, while a low $H_i$ value indicates small system travel times changes when agent $i$ chooses different routes. By ranking AV agents by this historical impact score $H_i$, we can reward a fraction of the most impactful AV agents with the marginal cost reward formulation. Selective marginal cost reward shaping therefore allows us to examine whether similar AV, HDV, and system travel times can be achieved with fewer counterfactual simulations.

\section{Results}

We begin by introducing the experimental setting in Section \ref{sec:experimental_setting}. Next, in Section \ref{sec:marginal_cost_performance}, we show that on two real-world traffic networks, selfish AV routing can disadvantage HDVs, whose travel times increase after AV introduction. We then demonstrate that rewarding all AVs with the marginal cost can mitigate the increase in HDV travel times and additionally improve AV travel times. Finally, in Section \ref{sec:selective_marginal_cost_performance}, we apply the marginal cost reward only to a fraction of the most impactful AVs, and show that rewarding just 40\% of AVs can achieve AV, HDV, and system travel times comparable to those obtained when marginal cost is applied to all AVs, while reducing the number of additional simulation runs by 40\%.

\begin{figure}[h]
    \centering
    \includegraphics[width=\linewidth]{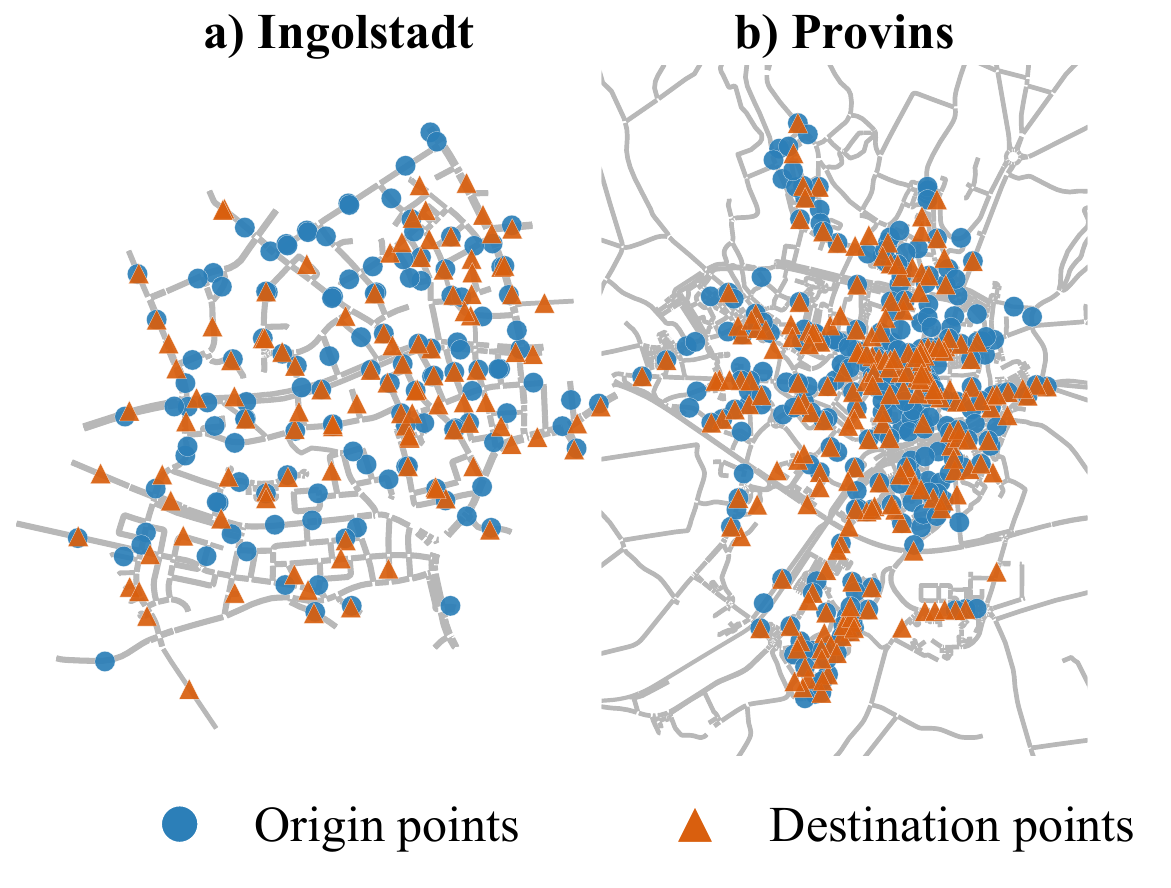}
    \caption{Traffic networks of Ingolstadt with 1035 agents (a) \cite{lobo2020intasingolstadttraffic}, and Provins (b) with 523 agents \cite{akman2025urburbanrouting}.}
    \label{fig:networks_topology}
\end{figure}

\begin{figure*}[h]
    \centering
    \includegraphics[width=\textwidth]{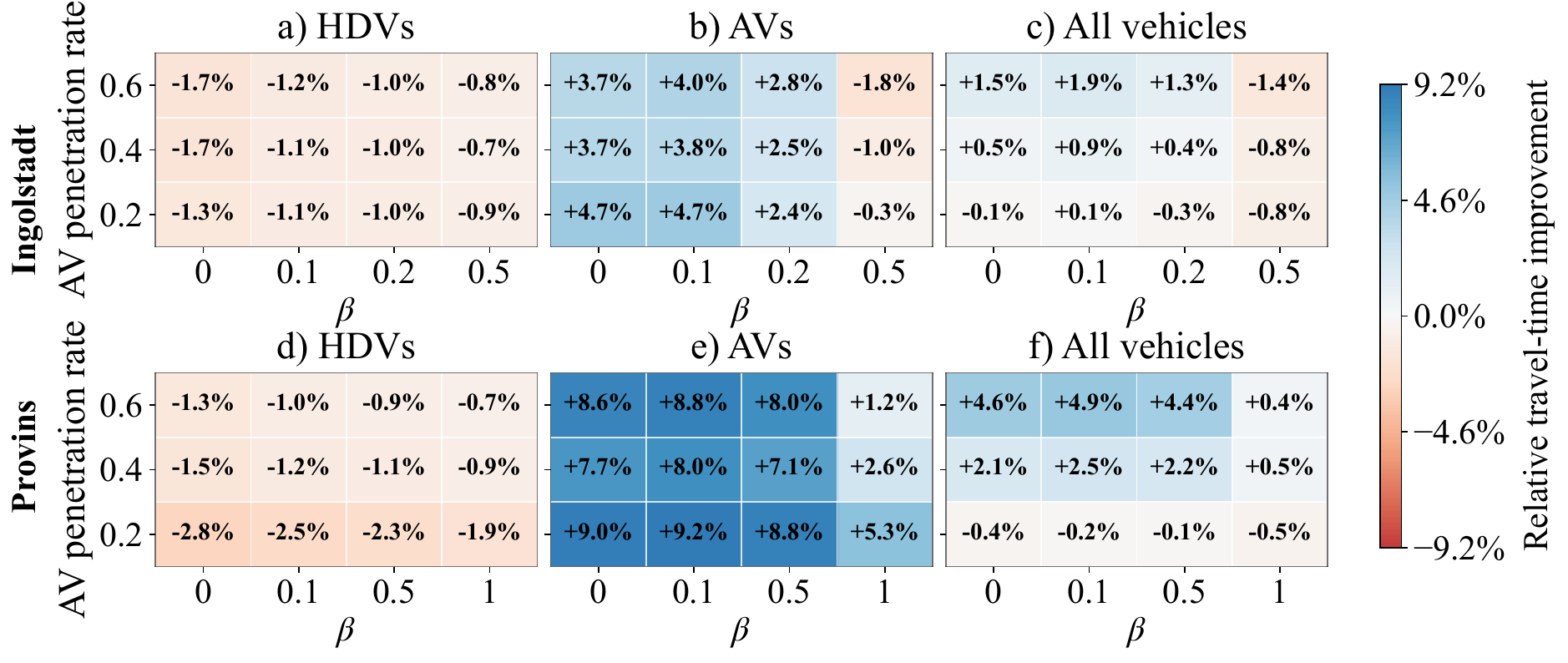}
    \caption{Average agent relative travel time change, computed as the difference between travel time before AV introduction and travel time during the testing phase, normalized by travel time before AV introduction. Negative values indicates the vehicle is disadvantaged after AV introduction, and a positive indicate the vehicle is better off after AV introduction. Values are averaged across three experimental seed replications, with standard deviations reported in the Supplementary Material.}    \label{fig:relative_travel_time_improvement}

\end{figure*}

\subsection{Experimental setting}
\label{sec:experimental_setting}

We assess the performance of our marginal cost reward formulation through experiments in the real-world Ingolstadt and Provins traffic networks, shown in Figure \ref{fig:networks_topology}. The Ingolstadt traffic network includes 850 nodes, 1426 links, 306 unique Origin-Destination (OD) pairs, and a demand of 1035 agents sourced from the InTAS traffic scenario \cite{lobo2020intasingolstadttraffic} and used in the RESCO benchmark \cite{ault2021reinforcement}. The Provins traffic network includes 2173 nodes, 4216 links, 517 unique OD pairs, and 523 traveling vehicles, sourced from the URB benchmark \cite{akman2025urburbanrouting}.

At the beginning of each experiment, the system contains only HDVs, which repeatedly choose routes among four options over 300 days of commute, as in \citep{jamroz2025social, akman2025routerlmultiagentreinforcementlearning, akman2025urburbanrouting}. On day 301, a specified fraction of HDVs is replaced by AVs that learn MARL routing policies using the REINFORCE algorithm \cite{10.1007/BF00992696, pmlr-v139-kim21g}, where each agent is an independent learner \cite{iql}. We use REINFORCE in our experiments, but the proposed reward formulation is not specific to this algorithm and can be incorporated into any MARL algorithm. We train the AVs for 10,000 training iterations or days of commute in both the Ingolstadt and Provins traffic networks, followed by a 100-day testing phase in which the learned policies are no longer updated. All results reported in Sections \ref{sec:marginal_cost_performance} and \ref{sec:selective_marginal_cost_performance} are computed over the 100-day testing phase. We report the REINFORCE algorithm hyperparameters in the Supplementary Material.

For selective marginal cost reward shaping, we identified the most impactful vehicles from the 800k joint action dataset $\mathcal{D}$. We sampled from enormous space of possible joint actions (for 1,000 vehicles where each vehicle can choose among four alternative routes, there are $4^{1000}$ possible joint actions) a diverse dataset $\mathcal{D}$ using several sampling methods, including random, balanced, Dirichlet \cite{song2024sampling}, Maximin distance sampling \cite{JOHNSON1990131}, and simplex-lattice design \cite{10.1111/j.2517-6161.1958.tb00299.x} (see Section \ref{sec:selective_marginal_cost}).

\subsection{Performance of the marginal cost}
\label{sec:marginal_cost_performance}

We demonstrate in Figure \ref{fig:relative_travel_time_improvement}(a, d) that in both the Ingolstadt and Provins real-world traffic networks, humans consistently experience longer travel times across all examined AV penetration rates when AVs selfishly minimize their individual travel times ($\beta=0$). The negative relative travel-time chage indicate that after AV introduction, HDV travel times increase. AVs, on the other hand, experience shorter travel times compared to those achieved by the human drivers they replaced at episode 300 (see Figure \ref{fig:relative_travel_time_improvement}(b, e) for $\beta=0)$. When marginal cost is incorporated into the AV reward ($\beta=0.1$ for Ingolstadt and Provins), the HDV relative travel time change increases relative to $\beta=0$, indicating that HDVs experience shorter travel times than under selfish AV routing. At the same time, AVs experience slightly shorter travel times on average, showing that \textbf{accounting for the system does not necessarily come at the expense of AV travel times}.

\begin{figure}[h]
    \centering
    \includegraphics[width=\linewidth]{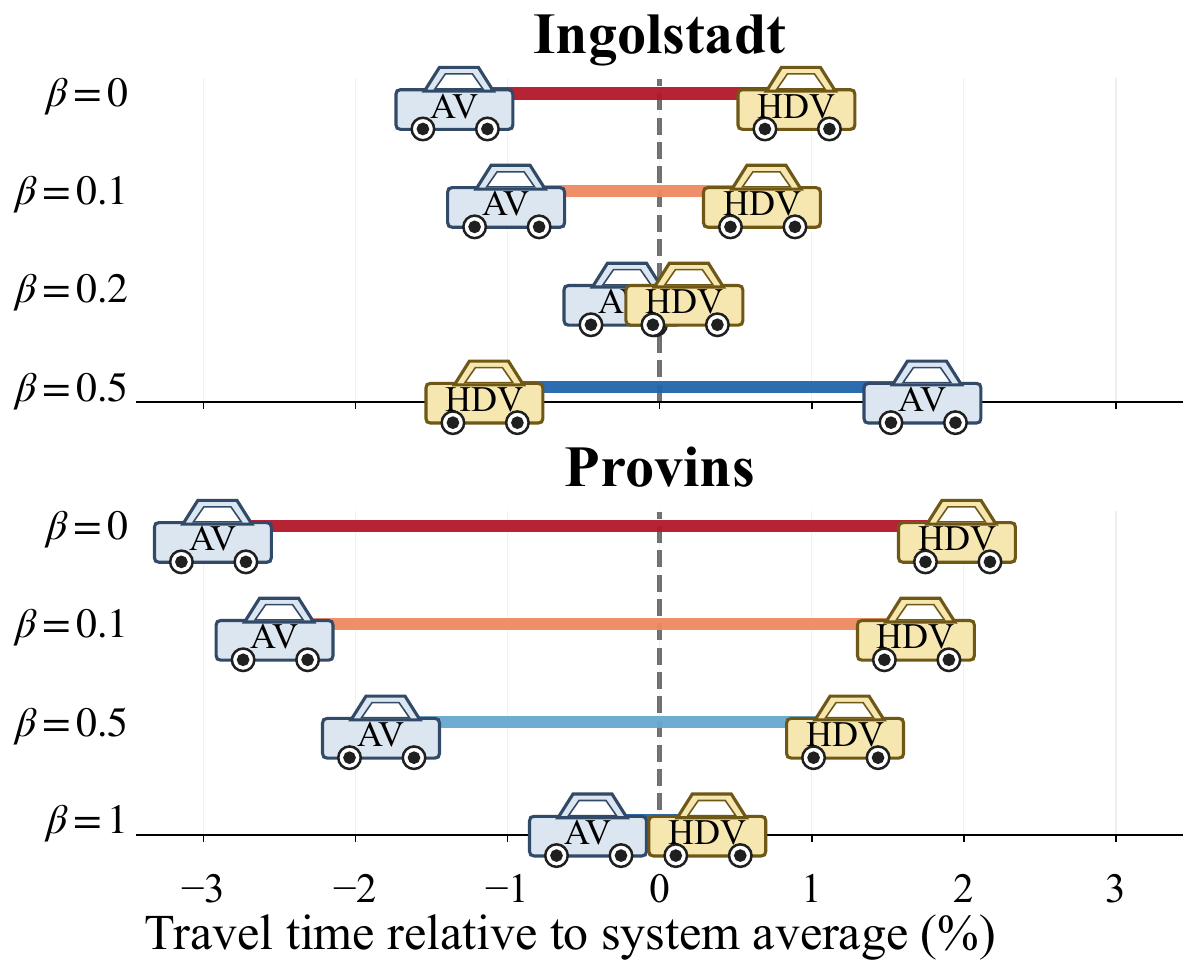}
    \caption{Change in average AV, and HDV travel times at 40\% AV penetration in Ingolstadt at various weights of social component $\beta$ in the reward. Negative values indicate shorter travel times compared to the system average.}
    \label{fig:human_av_tt_ratio}
\end{figure}

\begin{figure}[h]
    \centering
    \includegraphics[width=\linewidth]{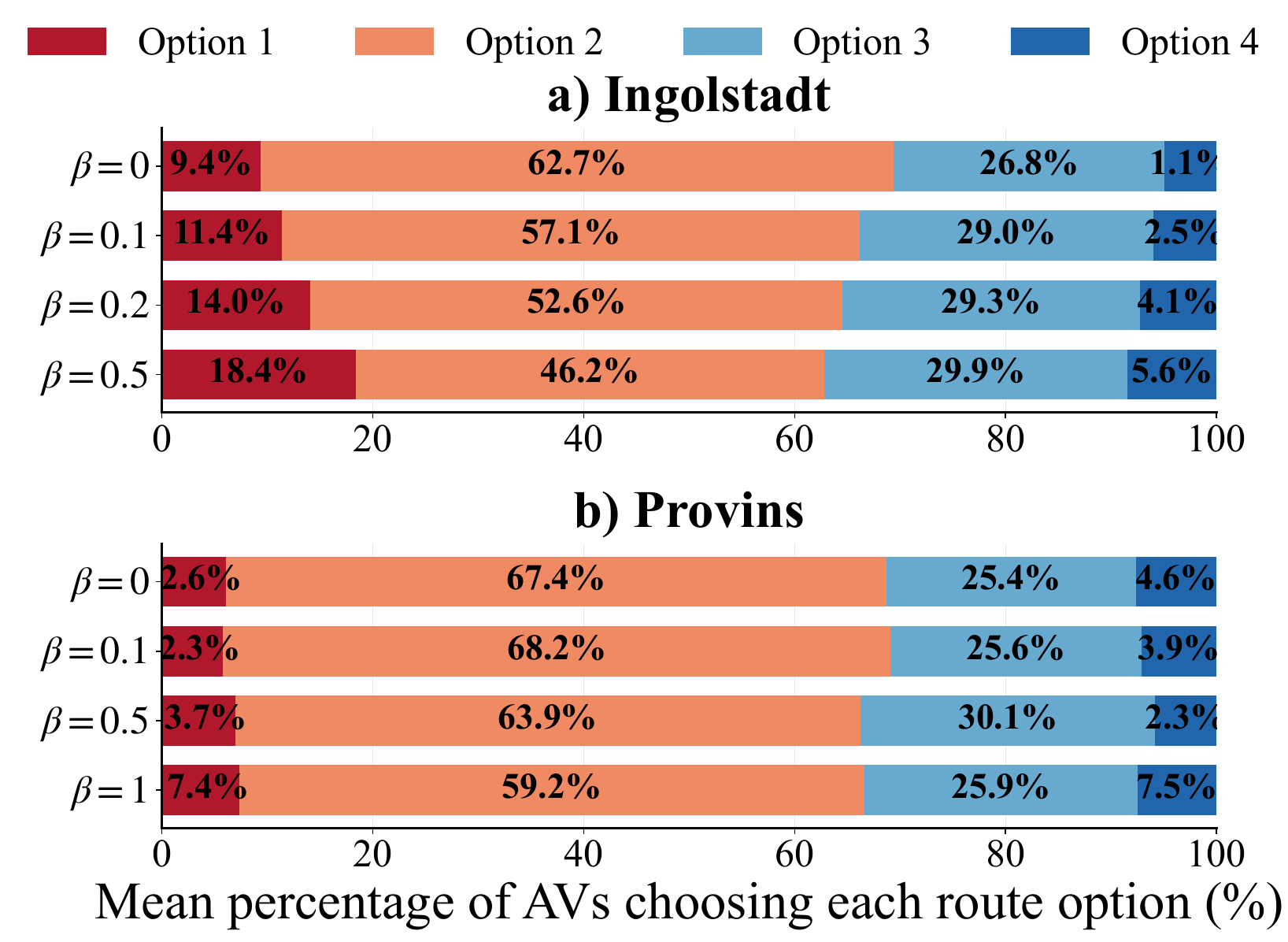}
    \caption{Mean percentage of AVs choosing routes from respective groups averaged over the 100 days of the testing phase on the Ingolstadt (a) and Provins (b) networks for a 40\% AV penetration rate.}
    \label{fig:route_choices}
\end{figure}

We then observe that as $\beta$ increases, e.g., $\beta=0.2, 0.5$ in Ingolstadt, and $\beta=0.5, 1$ in Provins, the relative travel-time change on HDVs increases, indicating that HDVs experience shorter travel times. However, this comes at the cost of longer AV travel times. Increasing $\beta$ beyond a certain value worsens both AV and overall system travel times (see Ingolstadt, $\beta=0.5$, Provins $\beta=1$).
\textbf{This indicates that placing too much weight
on the social component can disadvantage the system and individual AVs.}

Figure \ref{fig:human_av_tt_ratio} shows that for $\beta=0.1, 0.2$ in Ingolstadt, the gap between HDV and AV average travel times relative to the system average narrows, indicating that AVs, by accounting for their impact on the system, can reduce the travel-time gap between the two groups. At $\beta=0.5$, however, the HDV ratio decreases while the AV ratio increases, indicating that AVs placing greater weight on system performance can benefit HDVs relative to the system average while making AVs worse off. In Provins, increasing \(\beta\) from 0.1 to 1 progressively narrows the travel-time gap between AVs and HDVs, with the two groups approaching the system average at \(\beta=1\).

\begin{table*}[t]
    \centering
    \caption{Selective marginal cost reward shaping, specifically when 40\% or 50\% of the most impactful AVs are rewarded with the marginal cost, can achieve relative travel-time changes (see Figure \ref{fig:relative_travel_time_improvement}) close to those obtained when all AVs receive the marginal cost reward, while reducing the number of counterfactual simulations. The results are reported for the Ingolstadt traffic network with 40\% AV penetration. The AV subgroups column reports the relative travel-time change of selfish AVs (Selfish) and AVs receiving the marginal cost reward (MC). MC denotes marginal cost. Values are reported as mean $\pm$ standard deviation across experimental replications.}
    \label{tab:selective_mc_example}

    \setlength{\tabcolsep}{5pt}

    \begin{tabular*}{\textwidth}{@{\extracolsep{\fill}}lcccccc}
        \toprule

        Behavior
        & MC AVs
        & \makecell{Counterfactual\\simulations}
        & HDVs
        & AVs
        & AV subgroups
        & All \\

        \midrule

        \multicolumn{7}{l}{$\beta = 0.0$} \\

        Selfish
        & --
        & -
        & $-1.66 \pm 0.19$
        & $+3.74 \pm 0.75$
        & --
        & $+0.50 \pm 0.41$ \\

        \midrule

        \multicolumn{7}{l}{$\beta = 0.1$} \\

        MC
        & 100\%
        & 4,140k
        & $-1.06 \pm 0.17$
        & $+3.79 \pm 0.71$
        & \makecell{MC: $+3.79 \pm 0.71$}
        & $+0.88 \pm 0.39$ \\

        \arrayrulecolor{lightgrayrule}
        \specialrule{0.3pt}{2pt}{2pt}
        \arrayrulecolor{black}

        Selective MC
        & 50\%
        & 2,070k
        & $-1.17 \pm 0.18$
        & $+3.78 \pm 0.72$
        & \makecell{
            MC: $+3.51 \pm 0.74$ \\
            Selfish: $+4.05 \pm 0.70$
        }
        & $+0.81 \pm 0.39$ \\

        \arrayrulecolor{lightgrayrule}
        \specialrule{0.3pt}{2pt}{2pt}
        \arrayrulecolor{black}

        Selective MC
        & 40\%
        & 1,660k
        & $-1.25 \pm 0.17$
        & $+3.78 \pm 0.73$
        & \makecell{
            MC: $+3.30 \pm 0.75$ \\
            Selfish: $+4.10 \pm 0.72$
        }
        & $+0.76 \pm 0.38$ \\

        \arrayrulecolor{lightgrayrule}
        \specialrule{0.3pt}{2pt}{2pt}
        \arrayrulecolor{black}

        Selective MC
        & 10\%
        & 420k
        & $-1.50 \pm 0.19$
        & $+3.78 \pm 0.74$
        & \makecell{
            MC: $-0.18 \pm 0.70$ \\
            Selfish: $+4.21 \pm 0.74$
        }
        & $+0.61 \pm 0.40$ \\

        \arrayrulecolor{lightgrayrule}
        \specialrule{0.3pt}{2pt}{2pt}
        \arrayrulecolor{black}

        Selective MC (random)
        & 10\%
        & 420k
        & $-1.60 \pm 0.18$
        & $+3.75 \pm 0.72$
        & \makecell{
            MC: $+3.45 \pm 0.70$ \\
            Selfish: $+3.78 \pm 0.72$
        }
        & $+0.54 \pm 0.39$\\
        
        \bottomrule
    \end{tabular*}
\end{table*}

Differences in HDV, AV, and system travel times across values of \(\beta\) are accompanied by changes in AV route choices, as shown in Figure \ref{fig:route_choices}. As the weight on marginal cost increases, the mean percentage of AVs choosing routes from respective groups shifts from the one observed under selfish routing. In Ingolstadt, for example, the share of AVs choosing routes from the 2nd group decreases from \(\beta=0\) to \(\beta=0.5\), while the shares choosing Routes from groups 1, 3, and 4 increase. Here, Options 1-4 correspond to route options or groups rather than physical routes. In Ingolstadt, each of the 306 unique OD pairs has four candidate routes, which correspond to 1224 distinct route alternatives.

\subsection{Performance of the selective marginal cost}
\label{sec:selective_marginal_cost_performance}

We evaluate selective marginal cost reward shaping on the Ingolstadt traffic network at \(\beta=0.1\), where marginal cost improves travel times for AVs, HDVs, and the system compared with selfish AV routing. Our results show that applying this reward only to a subset of AVs yields relative travel-time changes (see Figure \ref{fig:relative_travel_time_improvement}) for HDVs, AVs, and the system that are close to those obtained when all AVs receive the marginal cost reward, while requiring fewer counterfactual simulations.

\begin{figure}[h]
    \centering
    \includegraphics[width=\linewidth]{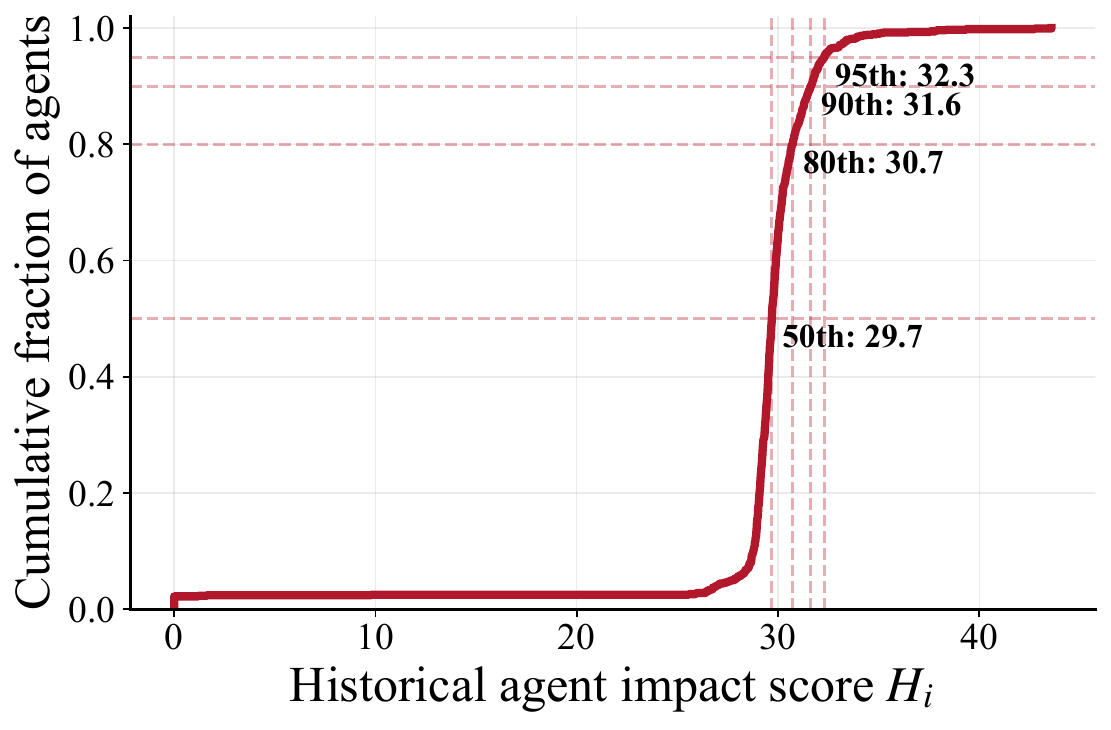}
    \caption{
    Empirical cumulative distribution of historical agent impact scores $H_i$ retrieved from dataset $\mathcal{D}$ of 800 thousand joint actions sampled in Ingolstadt. Dashed lines indicate the 50th, 80th, 90th, and
    95th percentiles.}
    \label{fig:impact_cdf}
\end{figure}

Figure \ref{fig:impact_cdf} shows the cumulative distribution of the historical impact ($H_i$ see Equation \ref{eq:historical_agent_impact}) of each of the 1035 vehicles on the Ingolstadt traffic network. Most vehicles have impact scores concentrated around 30. A smaller fraction of vehicles show lower or higher impact, indicating that vehicles differ in how strongly their route choices are associated with total system travel time.

Table \ref{tab:selective_mc_example} reports the relative travel-time change for HDVs, AVs, and all vehicles in the system, as shown in Figure \ref{fig:relative_travel_time_improvement}, when a fraction of the most impactful AVs receive the marginal cost reward. Specifically, when marginal cost is applied to the 50\% and 40\% most impactful AVs, HDV travel times improve by 0.5\% and 0.4\% compared to selfish AV routing, respectively, around 0.1\% and 0.2\% below the improvement obtained when marginal cost is applied to all AVs. We then tested a smaller fraction, specifically the 10\% most impactful AVs, and found that HDV travel times still improve by 0.16\% compared to selfish AV routing, although this is about 0.45\% below the case when marginal cost is applied to all AVs. Thus, \textbf{applying selective marginal cost to either 40\% or 50\% of the most impactful AVs could approximate the results of applying marginal cost to all AVs, retaining approximately 68\% and 82\% of the improvement in HDV relative travel-time change over selfish AV routing, respectively, while reducing the number of counterfactual simulations.}

To see if rewarding a fraction of the most impactful AVs based on $H_i$ is better than choosing a random fraction of AVs in the system. In the \textit{Selective MC (random)} row of Table \ref{tab:selective_mc_example}, we compare relative travel-time changes for HDVs, AVs, and all vehicles in the system when we apply marginal cost to a randomly selected fraction of AVs instead of the most impactful ones. We find a $0.1\%$ improvement in relative travel-time change for HDVs when we apply marginal cost to the 10\% most impactful AVs instead of a random 10\% of AVs in the system. 

For the Ingolstadt network at a 40\% AV penetration rate (around 414 AVs), 10k training iterations require about 4,140k counterfactual simulations. When we compute the marginal cost reward for 40\% of the AVs, this reduces to about 1,660k counterfactual simulations. We then need to account for the additional 800k joint-action samples of the dataset $\mathcal{D}$ used to identify the AVs with the largest system impact (see Figure \ref{fig:impact_cdf}), for a total of approximately 2,456k additional simulation runs. Thus, selective marginal cost reward shaping can reduce the additional simulation runs from 4 to 2.4 million, about a 40\% reduction, while achieving an average HDV relative travel-time change of about 0.2 percentage points below that obtained when marginal cost reward shaping is applied to all AVs.

In the \textit{AV subgroups} column of Table \ref{tab:selective_mc_example}, we compare relative travel-time changes of AVs receiving the selfish reward (\textit{Selfish}) and those receiving the marginal cost reward (\textit{MC}). When the 50\% or 40\% most impactful AVs receive the marginal cost, MC AVs experience smaller relative travel-time change than selfish AVs. This difference is more pronounced when marginal cost is applied to the 10\% most impactful AVs, where MC AVs experience a relative travel-time change of \(-0.18\%\), while selfish AVs experience a \(+4.21\%\) change. Thus, AVs that account for their impact on the system may be individually worse off, experiencing smaller relative travel time changes than selfish AVs.

\section{Conclusion}

AVs will increasingly operate in complex, capacitated urban traffic networks, where road users compete for limited road capacity, and each user's route choices can affect others' travel times. AVs may use a range of algorithms, with MARL being a promising one, to learn routing strategies that may differ from, and be more efficient than, those typically made by HDVs, potentially resulting in shorter AV travel times. Understanding how MARL AV routing can influence traffic performance is crucial to understanding how this emerging technology, AVs, may shape the performance of future urban transportation systems.

In this work, we study social AV route choice in mixed-autonomy traffic environments shared with HDVs, where AVs use MARL to learn routing policies. \textbf{To reduce the travel time costs AV routing imposes on other road users, we incorporate marginal cost into the AV reward.} Marginal cost is a counterfactual defined as the difference in travel times for all vehicles in the system when a given AV is present versus absent from the simulation. This formulation lets us include in MARL training the trade-off between minimizing an AV’s own travel time and reducing its impact on the system. 

Our results show that incorporating marginal cost into the AV reward can mitigate increases in HDV travel times while also reducing AV travel times relative to selfish AV routing, indicating that \textbf{AVs that account for their impact on the system can improve travel times for both HDVs, AVs, and the system}. However, placing greater weight on marginal cost can further improve HDV travel times at the expense of AVs and worsen total system performance. For example, at 40\% AV penetration in Ingolstadt, increasing \(\beta\) from 0 to 0.1 improves HDV, AV, and system travel time relative to the travel times before AVs were introduced by 0.6, 0.1, and 0.4 percentage points, respectively. Increasing \(\beta\) to 0.2 yields an additional 0.1 improvement in travel time change for HDVs but reduces the relative travel time change for AVs and the system by 1.3 and 0.5 percentage points, respectively, relative to \(\beta=0.1\).

Since computing the marginal cost requires an additional counterfactual simulation for each AV, we propose selective marginal cost reward shaping to reduce this computational burden. We find that assigning the marginal cost reward to a fraction of the most impactful AVs yields improvement in relative travel-time change for HDVs, AVs, and the system that are close to those obtained when all AVs receive the marginal cost reward. In the Ingolstadt traffic network, rewarding the 40\% most impactful AVs reduces the number of additional simulations by around 40\%, while the HDV relative travel time change is around 0.2 percentage points lower than when all AVs receive the marginal cost reward.

Overall, our findings show that AV routing algorithms can influence not only AV travel times but also how travel time costs are distributed across road users. AV routing policies that optimize only individual AV travel times may systematically disadvantage human drivers. By showing that marginal cost reward shaping can reduce these effects, our results highlight the importance of designing AV routing algorithms that account for AVs' impact on shared transportation systems.

\vspace{2em}

\bibliographystyle{plainnat}
\bibliography{sample}

\end{document}